\documentclass[conference,a4paper]{IEEEtran}
\IEEEoverridecommandlockouts
\usepackage{amsthm}
\usepackage{fancyhdr}
\usepackage{eso-pic}
\ifCLASSINFOpdf
\else
\fi
\usepackage{algpseudocode}
\usepackage[section]{algorithm}
\usepackage{amsmath}
\usepackage{amssymb}
\usepackage{mathtools}
\usepackage{graphicx}
\usepackage{subfigure}
\usepackage{caption}
\usepackage{cases}
\newcommand\bovermat[2]{%
  \makebox[0pt][l]{$\smash{\overbrace{\phantom{%
    \begin{matrix}#2\end{matrix}}}^{\text{#1}}}$}#2}

\begin{document}
\title{Fast and Optimal Power Control Games in Multiuser MIMO Networks}
\author{\IEEEauthorblockN{Peyman Siyari$^1$, 
Hassan Aghaeinia$^1$}
\IEEEauthorblockA{$^1$Department
of Electrical Engineering, Amirkabir University of Technology, Tehran,
Iran\\Email: {\tt psiyari@aut.ac.ir, aghaeini@aut.ac.ir} }
}
\maketitle
\begin{abstract}
In this paper, we analyze the problem of power control in a multiuser MIMO network, where the optimal linear precoder is employed in each user to achieve maximum point-to-point information rate. We design a distributed power control algorithm based on the concept of game theory and contractive functions that has a couple of advantages over the previous designs (e.g. more uniqueness probability of Nash equilibria and asynchronous implementation). Despite these improvements, the sum-rate of the users does not increase because the proposed algorithm can not lead the power control game to an efficient equilibrium point. We solve this issue by modifying our algorithm such that the game is led to the equilibrium that satisfies a particular criterion. This criterion can be chosen by the designer to achieve a certain optimality among the equilibria. Furthermore, we propose the inexact method that helps us to boost the convergence speed of our modified algorithms. Lastly, we show that pricing algorithms can also be a special case of our modified algorithms. Simulations show a noticeable improvement in the sum-rate when we modify our proposed algorithm. \end{abstract}

\IEEEpeerreviewmaketitle
\thispagestyle{fancy}
\section{Introduction}
\IEEEPARstart{I}{n} many wireless networks, such as ad-hoc networks or cognitive radio networks, there are multiple independent users that share a common communication environment. The interference channel is the most relevant mathematical model used to analyze the interaction of users in such networks. In order to calculate an inner bound for the capacity of the interference channel, the simplifying and pragmatic assumptions such as neglecting the use of interference cancellation techniques are common in the analysis of interference channels \cite{scutari1}. As these assumptions suggest, the analysis of the capacity of the interference channel simplifies to the analysis of power consumption of the users in network. Motivated by this fact, there are several studies about the power allocation of the users in an interference channel (see \cite{focc, yates, yu3} and references therein). 
In \cite{yates}, Yates et al. introduced the concept of standard functions that permits a general proof of the synchronous and totally asynchronous convergence of the iterative power control to a unique fixed point, minimizing the total transmit power for each user. Long after the work of Yates, Feyzmahdavian et al.\@ in \cite{feyz} introduced a slight and clever variation of the standard functions, namely as contractive functions to introduce a more unified explanation of distributed power control frameworks. Moreover, the concept of contractive interference functions was extended by the original authors to several cases of distributed power control.

The \lq\lq performance coupling\rq\rq among the users is the main challenge of using distributed approaches because the increase of one user's performance comes at the price of performance degradation of others.
One of the most applicable tools in distributed optimization is the use of game theoretical concepts. Motivated by this fact, we are interested in modeling the distributed power control in an interference channel as a non-cooperative game, where every user is a player that competes against the others by choosing the power allocation that maximizes its own information rate. A Nash equilibrium is achieved when no player can unilaterally increase its own information rate given the current strategies of the others \cite{funden}. Analysis of power control in interference channel using non-cooperative game theory was first done in \cite{yu3}, where Yu, et al.\@ solved the power control problem in Digital Subscriber Line (DSL) systems by modeling the system as a gaussian frequency-selective interference channel and performing Iterative Water-Filling Algorithm (IWFA). Since then, a number of studies have been done to extend this original idea, and finally, the studies of Scutari et al.\@ brought a unified view of distributed IWFA in many scenarios such as multiuser MIMO and cognitive radio systems (see \cite{sc1, sc3}).

In this paper, distributed power control in multiuser MIMO systems, is investigated with the help of game theory and contractive interference functions. 
Our approach is compared with the state-of-the-art method used in \cite{sc3}, where the authors give a different interpretation of water-filling by using it as a contraction mapping. We show that with the use of the same amount of signaling as in the previous papers, the uniqueness of Nash equilibria is more probable in our approach. Furthermore, we show that our distributed power control algorithm is flexible enough to deal with various practical limitations of communications networks such as communication delays.

Although our algorithm has a number of advantages over the previous studies, the sum-rate maximization is not achieved. This is because of the fact that the equilibrium points achieved with non-cooperative games may not be Pareto optimal points. This issue may steer the interest away from using our algorithm in practice. Therefore, the issue of performance coupling among the users must be taken into account. To overcome this issue, we seek for a different definition of non-cooperative power control games using the concepts of variational inequality. With this definition, we modify our proposed algorithm such that the equilibrium point achieved in the power control game is the one that satisfies a certain criterion for us. The optimality criterion can be chosen arbitrarily in order to achieve different optimality conditions. When we choose sum-rate maximization as our criterion, the power control game achieves more efficient equilibrium points in terms of sum-rate. We also propose an inexact method to reduce the number of iteration for our modified algorithms, which makes our algorithms more suitable for practical implementation.

The rest of this paper is organized as follows. In section II, the general system model is described. In section III, the power control problem is formulated as a non-cooperative game, and the elements of the game are determined. In section IV, a brief introduction of contraction mappings and contractive functions is presented. After that, the analysis of uniqueness of Nash equilibrium is done, and the condition that guarantees the uniqueness of NE in power control game is established. Next, we discuss the design of distributed power control algorithms in section V. In section VI, we show the strength and the weakness of our algorithm, then we define the concept of variational inequality to make foundations for our further analysis of the power control game, In section VII, we try to find a solution to eliminate the weakness of our algorithm by designing new modified versions of it. In Section VIII, the computer simulations verify our theoretical analysis in the previous section, and it is shown that we were able to overcome the weakness of our algorithm. Finally, section IX concludes the paper.
\section{System Model}
In this paper, we employ a multiuser MIMO channel as our system model. There are Q transmit-receive pairs, and each pair is a user. The $q$th transmitter has $N_{T_q}$ transmit antennas and the $q$th receiver has $N_{R_q}$ receive antennas. The received signal at each receiver interfered by the other $Q-1$ transmitters. As we do not allow any multiuser encoding or coordination among the users, the interference at each receiver is treated as an additive component. Hence, the received signal at the $q$th receiver (i.e, $y_q$) is
\begin{align}
\label{sysmodel}
y_q = H_{qq}x_q+\sum_{r \neq q}H_{rq}x_r+n_q,
~~\forall q\in\{1,...,Q\}
\end{align}
where $x_q \in C^{N_{T_q}}$ is the complex transmitted signal from the $q$th transmitter, $y_q \in C^{N_{R_q}}$, $H_{rq} \in C^{N_{R_q}\times N_{T_r}}$ is the complex channel matrix between the $r$th transmitter and $q$th receiver, and $n_q$ is the additive white Gaussian noise (AWGN) vector with the covariance matrix $R_{n_q} = N_{0_q}I_{N_{R_q}\times N_{R_q}}$ where the scalar value $N_{0_q}$ is the power of the noise and $I$ is the identity matrix. Hence, the vector $\sum_{r \neq q}H_{rq}x_r$ is the multiuser interference (MUI) at the $q$th receiver. Given $\mathcal{E}\left\{ x_qx_q^H\right\}$ as the covariance matrix of the vector $x_q$, the power constraint for each transmitter is
\begin{equation}
\label{powerconstraint}
\mathcal{E}\left\{ ||x_q||_2\right\} = Tr\left(\mathcal{E}\left\{ x_qx_q^H\right\}\right)\le P_q,
\end{equation}
where $||.||_2$ is the second vector norm, $Tr(.)$ is the trace operator, and $P_q$ is a scalar value that represents the total amount of power that a transmitter can distribute between its antennas.  Assuming that the vector $p_q = [p_q^1, ...,p_q^{N_{T_q}} ]^T = diag\left(\mathcal{E}\left\{ x_qx_q^H\right\}\right)$ indicates the power of the transmitted signal from the $q$th transmitter, the power constraint for the $q$th user can alternatively be shown as:
\begin{equation}
\sum_{i = 1}^{N_{T_q}}p_q^i \le P_q.
\end{equation}
\begin{figure}
\centerline{
\includegraphics[scale = 0.3]{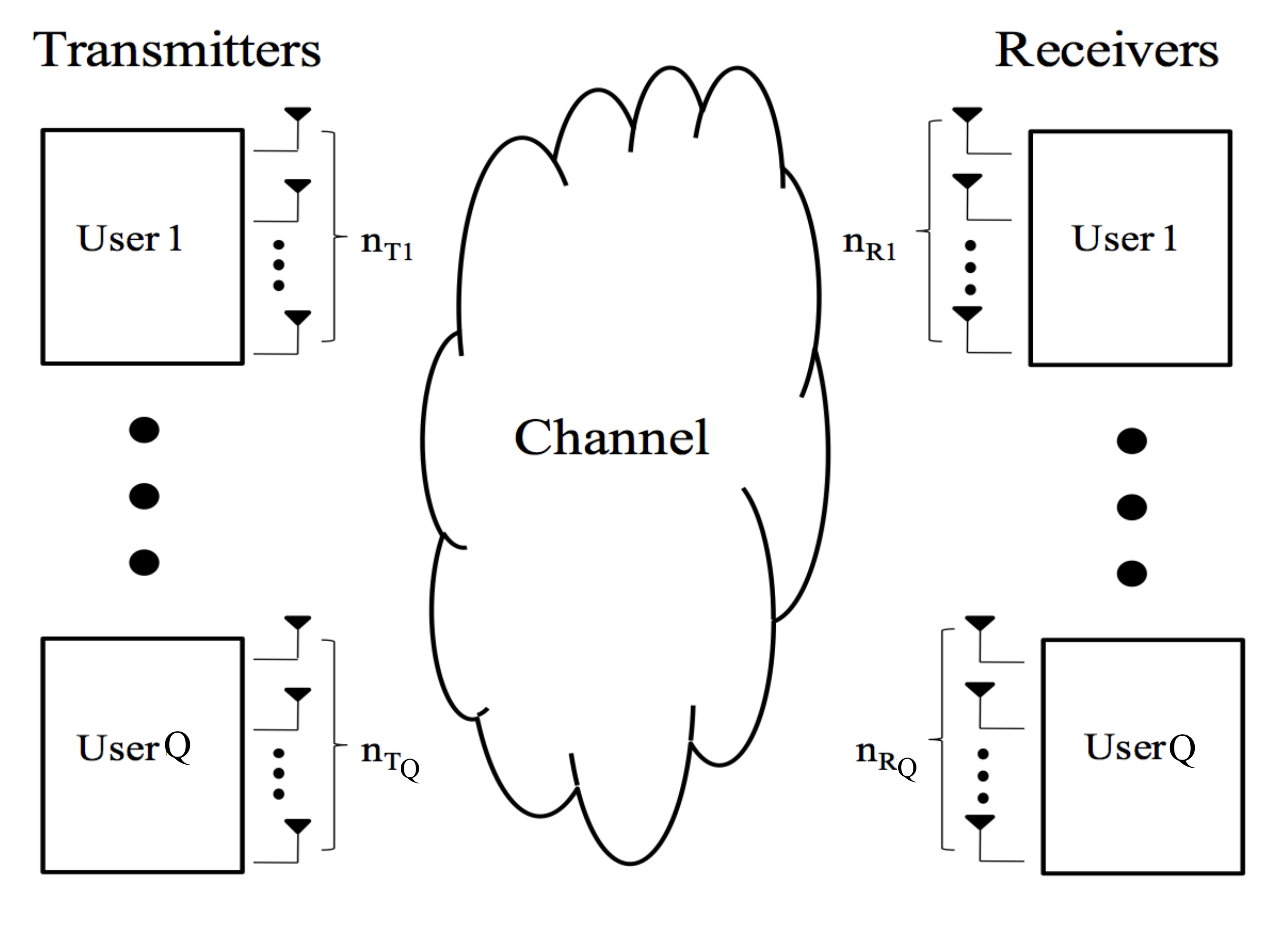}
}
\caption{\small System Model}
\end{figure}
\section{Problem Formulation}
\subsubsection*{Using Channel State Information (CSI)}
We assume that each transmitter knows the channel between itself and its receiver pair \footnote{This assumption imposes no overhead more than the previous related work (compare with \cite{sc3}).}. It can be shown that the knowledge of CSI (in i.i.d channels) at the transmitter can significantly improve the capacity of channel \cite{paulraj}. The knowledge of CSI can be exploited by designing an appropriate precoder, and the capacity-achieving scheme is to use singular-value decomposition to diagonalize the channel between a transmit-receive pair. The singular-value decomposition of $H_{qq}$ yields
\begin{equation}
H_{qq} = U_q \Sigma_q V_q^H,
\end{equation}
where $U_q$ and $V_q$ are unitary matrices and $\Sigma_q$ is the diagonal matrix of singular values of $H_{qq}$. We assume the vector $diag(\Sigma_{q}) = \sigma_q = [\sigma_q^1, ..., \sigma_q^{\nu_{q}}]^T$ (where $\nu_q = min(N_{T_q},N_{R_q})$) is the vector of the singular values of $H_{qq}$. Using $V_q$ as the precoder matrix and $U_q^H$ as the decoding matrix, the received signal at the $q$th receiver would be
\begin{equation}
\label{svd}
\tilde{y}_q = \Sigma_{q}x_q+\sum_{r \neq q}U_q^HH_{rq}V_r x_r+U_q^Hn_q,
\end{equation}
where $\tilde{y}_q = U_q^Hy_q$. Since $U$ and $V$ are invertible, knowing $\tilde{y}_q$ is the same as knowing $y_q$. Furthermore, the noise vector's statistics do not change when multiplied by $U_q^H$, since $U_q$ is a unitary matrix and $n_q$ has independent Gaussian elements \cite[Chapter 3]{duman}. We take $\tilde{H}_{rq} = U_q^HH_{rq}V_r$ as an $N_{R_q} \times N_{T_r}$ matrix. The element in the $i$th row and $j$th column of $\tilde{H}_{rq}$ is shown by the operator $\left[\tilde{H}_{rq}\right]_{i,j}$. Assuming no correlation between the transmit and receive antennas, the capacity of the MIMO channel between the $q$th transmit-receive pair would be:
\begin{equation}
\label{rate}
R_q = \sum_{i=1}^{\nu_q} \log\left(1+\frac{p^i_q}{c^i_q}\right),
\end{equation}
where $c^i_q$ --the normalized interference plus noise of the $q$th user-- is written as
\begin{equation}
c^i_q = \sum_{r \neq q}\left\{\sum_{j = 1}^{{N_T}_r}\frac{\left|\left[\tilde{H}_{rq}\right]_{i,j}\right|^2p^j_r}{|\sigma_q^i|^2}\right\}+\frac{N_{0_q}}{|\sigma_q^i|^2},~~i\in\{1,...,\nu_q \}.
\end{equation}
\subsubsection*{Game Formulation}
We assume that each user myopically chooses the best strategy for itself, so we formulate this scenario as a non-cooperative game, in which the best strategy of each user is
\begin{align}
\label{optimisation}
\max_{p_q}~R_q
\\
\text{s.t.}~\sum_{i = 1}^{N_{T_q}}p^i_q \le P_q
\notag.
\end{align}
Therefore, in this game, the utility function of each player (or user) is its point-to-potint information rate and the strategy of each player is choosing the best power allocation according to its power constraint (i.e., strategy set) to maximize its rate. The maximum information rate is achieved by performing water-filling which can be shown as
\begin{equation}
\label{water}
{p^i_q}^* = \left(\mu_q-c^i_q\right)^+,
\end{equation}
where $(...)^+ = max(...~,~0)$ and $\mu_q$ is a positive value that satisfies the power constraint.

The existence of Nash equilibrium in this game can be proven by showing that the strategy set of each user is non-empty, compact and convex subset of some Euclidean real space $\mathcal R^{N_{T_q}}$, and the utility function of each user is a continuous and quasi-concave function of its own power vector. The detailed proof for the existence of NE is straight-forward and skipped for brevity. The uniqueness of equilibrium is investigated in the next section.
\section{Uniqueness of Nash Equilibrium}
So far, we have formulated a non-cooperative power control game in a multi-user MIMO system, in which every transmitter performs water-filling in allocating the power between its antennas. In this section, we introduce a new characterization of MIMO power control. First, we review fundamentals of fixed-point theory as the basis of our analyses.
\subsubsection*{Fixed-Point Theory}
Consider the following iteration:
\begin{equation}
\label{fix}
x(k+1) = T\left(x(k)\right),~k=1,2,...,
\end{equation}
where $T$ is a mapping from a subset $X$ of $\mathcal R^K$ to itself, and $k$ indicates the index of iterations. If T is continuous and
\begin{equation}
||T(x)-T(y)|| \le c||x-y||~,~\forall x,y \in X
\end{equation}
where $||.||$ is some norm in $X$ and $c \in [0,1)$, then the mapping $T$ is a contraction mapping with $c$ as the contraction modulus, and sequence $\left\{x(n)\right\}$ generated by the iterations in \eqref{fix} converges to the fixed point $x^*$ \cite[Chapter 13]{feyz13}.

\subsubsection*{Contractive Functions}
A function $I: \mathcal R^K_+ \rightarrow \mathcal R^K_+$ (where $\mathcal{R}^K_+ = \{p|p>0, p\in \mathcal{R}\}$) is said to be contractive if it, for all $p \ge 0$ satisfies \cite{feyz}:
\begin{itemize}
\item{}
Positivity: $I(p)\ge 0$
\item{}
Monotonicity: if $p \ge p'$, then $I(p) \ge I(p')$
\item{}
Contractivity: There exists a constant $c \in [0,1)$ --namely as the contractivity modulus-- and a vector $v > 0$ such that $I(p+\alpha v) \le I(p) + c\alpha v,~~\forall \alpha > 0$.
\end{itemize}
The following results have been deriven about contractive functions \cite{feyz}:
\newtheorem{proposition}{\bf{Proposition}}
\begin{proposition}
If a function $I: \mathcal R^K_+ \rightarrow \mathcal R^K_+$ is contractive with $c\in[0,1)$ as the contractivity modulus, then
\begin{enumerate}
\item{}
It is a contraction mapping with maximum norm and $c\in [0,1)$ as the contraction modulus, then it has a unique fixed point $p^*$, and the same iteration as in \eqref{fix} for $I(p)$ will converge to $p^*$.
\item{}
The following function is also contractive with the same modulus $c \in [0,1)$:
\begin{equation}
I^q(p) = max\left\{p_{min}, min\{p_{max}, I(p)\}\right\},
\end{equation}
where $p_{min}$ and $p_{max}$ can be the minumum and miaximum constraints on the power allocation, respectively.
\end{enumerate}
\hfill $\square$
\end{proposition}
With these results, we are now ready to present a new theorem on the convergence of distributed power control in our scenario.
\newtheorem{theorem}{\bf{Theorem}}
\begin{theorem}
The MIMO power control game defined in the previous section with the players' strategies written as \eqref{water} has a unique Nash equilibrium if $\forall (i,q)$ we have
\begin{equation}
\label{th1}
\max_{(i,q)}\sum_{r\neq q}\sum_{j=1}^{N_{T_r}} \frac{\left|\left[\tilde{H}_{rq}\right]_{i,j}\right|^2}{|\sigma_q^i|^2}<1,
\end{equation}
or alternatively
\begin{equation}
\label{th2}
\max_{(i,q)}\sum_{r\neq q}\sum_{j=1}^{N_{T_r}} \frac{\left|\left[\tilde{H}_{qr}\right]_{i,j}\right|^2}{|\sigma_q^i|^2}<1.
\end{equation}
\begin{IEEEproof}
See Appendix A.
\end{IEEEproof}
\end{theorem}
\section{Algorithm Design}
\label{design}
In the previous section, it was proved that the distributed iterative waterfilling that each user performs is a conrtactive function (, and therefore a contraction mapping with maximum norm). The immediate result of this proof is that the iterative water-filling algorithm can be done totally asynchronously in the sense of \cite{feyz13}. Let $T_q,~\forall q \in \{1,2,...,q\}$ be the set of times (or iteration numbers) when $q$th user updates its power allocation. While updating the power allocation,  it is possible that an outdated power radiated by the interfering links is used by a user in the calculation total interference, so the power allocation may not be according to the recent changes in the network. This usually happens due to the communication delays that may occur in the network. To model this issue in the network, we assume that $\theta^{q^{(n)}} = \{\theta_1^{q^{(n)}}, ...,\theta_Q^{q^{(n)}}\}$ as the set of most recent times that the information from each user is received by the $q$th user at the $n$th iteration. Therefore, at each iteration $n \in T_q$ the $q$th transmitter performs waterfilling based on the $\theta^{q^{(n)}}$ that is available at the $n$th iteration. With these definitions, we are ready to present our asynchronous power control algorithm for the iteration number $n$ which is as follows:

\begin{algorithm}[H]
\caption{Asynchronous Iterative Water-filling}
\label{alg}
Set~$p_q(0)$ such that 
$\sum_{i = 1}^{N_{T_q}}p^i_q \le P_q$
\begin{algorithmic}[1]
\For{n=1 to $it_{max}$}
\Repeat
~~~$\forall (i,q)\in\{1,...,\nu_q\}\times\{1,...,Q\}$
\State
$
{{p^i_q}^{(n)}} =\left\{
\begin{array}{c l}     
\left({\mu_q}^{(n-1)}-{c^i_q}^{(n-1)}\right)^+ & \text{if~}n\in T_q\\
    {{p^i_q}^{(n-1)}} & \text{otherwise}
\end{array}\right.$,
\Until{Convergence}
\EndFor
\end{algorithmic}
\end{algorithm}
In the above algorithm, $it_{max}$ denotes the maximum number of iterations, and the fourth step can be replaced with a particular termination criterion. Furthermore, in the term ${c^i_q}^{(n)}$, the most recent power updates of other users (according to $\theta^{q^{(n)}}$) are available. The special cases of asynchronous implementation include synchronous implementation (Jacobi algorithm \cite{feyz13}) and sequential implementation (Gauss-Seidel algorithm \cite{feyz13}), that are used in the previous algorithms \cite{yu3, sc1}. For the Jacobi Algorithm we have \begin{align}
T_q = \{1, 2, ..., it_{max}\}
,\ 
\theta^{q^{(n)}} = \{n,...,n\}
\end{align}
which means at each iteration all of the users simultaneously update their power allocation, and for Gauss-Seidel we have 
\begin{align}
T_q = \{q, q+Q, q+2Q, ..., q+\left(\frac{it_{max}}{Q}-1\right)Q\} ~~~~~~~~~~~~
\notag
\\
\theta^{q^{(n)}} = \left\{ \begin{array}{c l}     
\{n-(q-1), ..., n-1\} & for j=1:q-1
\\
\{n, n-(Q-1), ..., n-q\} & for j=q:Q
\end{array}\right.
\end{align}
which means at each iteration only one user updates its power allocation and all the other users do not update, and this procedure sequentially continues between the users. It should be noted that due to its sequential nature, the Gauss-Seidel algorithm may require a proper scheduler in the network. Depending on the limitations of the network in terms of communication delay, one can use different update patterns, which shows the flexibility of asynchronous power control algorithm in practical situations\footnote{In the simulation part, we only use synchornous power control (i.e. Jacobi method in the sense of \cite{feyz13}) as it converges quite fast.}. In sum, as long as the transmitters are intended to update their power allocations (i.e., the set $T_q$ has infinite elements \cite[Chapter 6]{feyz13}) and all the transmitters eventually become aware of other transmitters' interference (i.e., $\lim_{n\rightarrow \infty}{\theta^q}^{(n)} = \infty$ \cite[Chapter 6]{feyz13}), our power control algorithm converges under any update pattern. This result is valuable as it is applicable to other approaches that only prove the convergence of power control under Gauss-Seidel algorithm and the convergence of Jacobi algorithm is proved by conducting simulations with no theoretical analysis (see \cite{diep2}). Moreover, the convergence under asynchronous iteration mitigates the need for scheduler (which is the disadvantage of Gauss-Seidel algorithm) and accurate synchronization (which is the disadvantage of Jacobi algorithm).
\section{Performance Analysis and Modification}
In this section, we evaluate the analyses done in the previous sections and compare them to the other reference's work. These comparisons reveal the strength and the weakness of our algorithm.
\subsection{Performance Analysis}
\label{perf.anlsys}
 We simulated a network comprised of four MIMO links that share the same band. 
 The wireless channel is comprised of an i.i.d complex Gaussian flat-fading with zero mean and unit variance for the small-scale fading, 
and exponential path-loss for the large-scale fading. The additive noise is set to i.i.d complex Gaussian with zero mean and unit variance. We assume that a given scenario $N_T \times N_R$ indicates $N_T$ transmit antennas and $N_R$ receive antennas for all of the users, meaning that $N_{T_q} = N_T~\&~N_{R_q} = N_R,~\forall q \in\{1,...,Q\}$). All of the users have the same total power budget $P_q = 10 dB,~\forall q$. The distance between the $q$th transmit-receive pair is shown as $d_{qq}$ and the distance between the $r$th transmitter and the $q$th receiver $d_{rq}$. Lastly, the path-loss exponent is set to $\gamma = 2.5$.

In this simulation, we want to quantify how adequately our derived conditions in \eqref{th1} and \eqref{th2} can predict the convergence of the power control game in practice. Fig. 2 shows the probability of uniqueness of NE as a function of $d_{rq}$, where $d_{qq} = 15$. Each point on the curves, is the result of the total number of times that a particular criterion is satisfied (and the game also converges to the unique NE) divided by the total number of channel realizations. 
There are 1000 total channel realizations for each point of the curves. Each realization is done within the maximum of 100 iterations of Jacobi method. The termination criterion is that if the normalized difference between the two consecutive iterations becomes smaller than $10^{-4}$ the algorithm ends. Three criteria are used for each antenna configuration. The curves with the stars indicate whether the condition in \cite[Condition 7]{sc3} is satisfied.
\begin{figure}
\label{fig2}
\centerline{
\includegraphics[scale = 0.57, trim= 370mm 90mm 350mm 100mm]{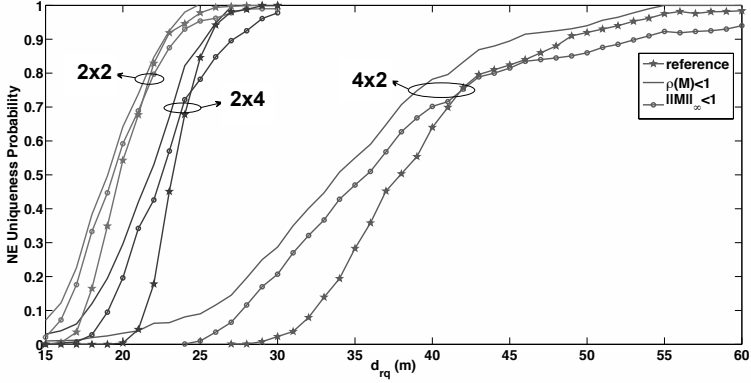}
}
\caption{\small Comparison of probability of uniqueness of NE for different number of transmit/receive antennas w.r.t. interfering distance}
\end{figure}
 The curves with circles indicate the probability of uniqueness according to the satisfaction of \eqref{th1} or \eqref{th2}. Finally the solid curves indicate the convergence to the unique NE if $\rho(M) <1$ is satisfied. First of all, both of the convergence criteria we derived predict the uniqueness of NE more than the condition derived in \cite{sc3} for small distances because the precoding employed in our approach uses the channel information more efficiently. Furthermore, the condition $\rho(M) < 1$ predicts the uniqueness more than the condition in \eqref{th1}. As it was said, the conditions in \eqref{th1} and \eqref{th2} are derived by setting the weighting vector $v = 1$ in $||M||^v_\infty$. By setting $v = 1$, the probability of uniqueness of NE becomes less than the reference curve as the interfering distances increase\footnote{Note that the conditions derived in \cite{sc3} are not practical as well. The practical conditions derived in \cite{sc3} have worse performance loss.}. In fact, when the condition $\rho(M) < 1$ is satisfied,  the vector $v$ can have any value, but setting $v = 1$ leads to more conservative conditions that ignores some situations wherein the vector $v$ has a value other than $v = 1$. However, setting $v = 1$ leads to more practical conditions that are suitable for network designing. Lastly, for the case $N_T = 4~\&~N_R = 2$ (which is shown as $4 \times 2$ in Fig. 2), our criteria have the same trend as the criterion in \cite{sc3} does, making our proof for the case $N_T > N_R$ reasonable.
\subsection{Alternative Definition of the Power Control Game}
Although there are some improvements in our proposed algorithm, the sum-rate of the users has not been improved yet, which makes our algorithm (and most of non-cooperative power control games) less favorable for practical implementations. This is due to the fact that the Nash equilibria of the game is in general not Pareto optimal, then the selfish maximization of rate cannot be guaranteed to achieve global optimality. In order to modify our algorithm, we need another definition of power control game. This definition is based on the concept of variational inequality. In the following, we define this concept and its application in non-cooperative game theory.
\newtheorem{definition}{\bf{Definition}}
\begin{definition}
Let $F: \mathcal{Q} \rightarrow \mathcal{R}^n$ be a real-valued vector function, where $\mathcal{Q}$ is a nonempty, closed, and convex set. The variational inequality $VI(\mathcal{Q},F)$ is the problem of finding a vector $x^*$ such that the following inequality is satisfied:
\begin{equation}
(x-x^*)^TF(x^*)\ge0,~~\forall x\in\mathcal{Q}.
\end{equation}
\end{definition}
This problem is a generalization of minimum principle in convex optimization. For a convex function $f(x)$, the point $x^*$ is the minimum of $f$ if the following inequality is satisfied:
\begin{equation}
\label{optcond}
(x-x^*)^T\nabla f(x^*)\ge0,~~\forall x\in\mathcal{Q}.
\end{equation}
The relation between variational inequality and game theory is summarized in the following theorem:
\begin{theorem}
Consider $Q$ players in a non-cooperative game with convex utility functions $f_q(x)~\forall q$, where $x$ is a vector comprised of other vectors (i.e., $x = [x_1,x_2,...,x_Q]^T$). Assuming closed and convex solution sets for $f_q(x)~\forall q$, the vector $x^*$ is the Nash equilibrium of the game if for the function $F(x) = [\nabla_{x_1}f_1(x),\nabla_{x_2}f_2(x),...,\nabla_{x_Q}f_Q(x)]^T$we have \cite{pang}:
\begin{equation}
(x-x^*)^TF(x^*)\ge0,~~\forall x\in\mathcal{Q}.
\end{equation}
\qed
\end{theorem}
Next, we state an important property in variational inequality, which will be used later in this paper.
\begin{theorem}
For a mapping $F: \mathcal{Q} \rightarrow \mathcal{R}^n$, that is continuously differentiable on $\mathcal{Q}$, and has the Jacobian matrix denoted by $\mathcal{J}$, it hold that $\forall x \in \mathcal{Q}$ \cite{pang}:
\begin{itemize}
\item{}
$F(x)$ is monotone on $\mathcal{Q}$ iff $\mathcal{J}(x)$ is positive semidefinite. Therefore, the problem $VI(\mathcal{Q}, F)$ has a (possibly empty) convex solution set.
\item{}
$F(x)$ is strictly monotone on $\mathcal{Q}$ if $\mathcal{J}(x)$ is positive definite. Therefore, the problem $VI(\mathcal{Q}, F)$ has at most one solution.
\item{}
$F(x)$ is strongly monotone on $\mathcal{Q}$ iff $(\mathcal{J}-c_{sm}I)$ is positive semidefinite ($c_{sm}$ is the strong monotonicity modulus). Therefore, the problem $VI(\mathcal{Q}, F)$ has a unique solution.
\end{itemize}
\qed
\end{theorem}

In the following, we try to write the power control game as a variational inequality problem. To do so, we address an important property of KKT conditions.
\begin{proposition}
Consider the optimization problem \eqref{optimisation}. Replacing the negative of the objective function, we have the following KKT conditions:
\begin{align}
\label{kkt1}
0\le p^i_q~~\perp~~-\frac{1}{\left(c^i_q+p^i_q\right)}-\lambda^i_q\ge 0,
\\
0\le \lambda^i_q~~\perp~~\sum_{i = 1}^{\nu_q}p^i_q-P_q = 0.~~~~~~~
\notag
\end{align}
With the additional constraints\footnote{These constraints impose a limit on the maximum power that each antenna can transmit, which are reasonable in practice} $p^i_q \le P^i_q$ and $P_q < \sum_iP^i_q$, the KKT conditions are equivalent to \cite{luo}:
\begin{align}
0\le p^i_q~~\perp~~\left(c^i_q+p^i_q\right)-\lambda^i_q-u^i_q\ge 0,
\\
0\le \lambda^i_q~~\perp~~\sum_{i = 1}^{\nu_q}p^i_q-P_q = 0,~~~~~~~~~
\notag
\\
0\le u^i_q~~\perp~~p^i_q-P^i_q \le 0.~~~~~~~~~~~~~~
\notag
\end{align}
Therefore, the optimization problem \eqref{optimisation} can be written as
\begin{align}
\label{optequi}
\min_{p_q}\frac{1}{2}\sum_{i=1}^{\nu_q}||\left(c^i_q+p^i_q\right)||^2
\\
\notag
\text{s.t.}~\sum_{i = 1}^{\nu_q}p^i_q<P_q.
\end{align}
\qed
\end{proposition}
It is worth mentioning that the term $-\frac{1}{\left(c^i_q+p^i_q\right)}$ in the first line of \eqref{kkt1} is actually $-\nabla_{p_q}R_q$. Therefore, setting $F(p) = [-\nabla_{p_1}R_1, ..., -\nabla_{p_Q}R_Q]^T$ the problem $VI(\mathcal{Q}, F)$ can be demonstrated. Using the above proposition, we can set $F(p) = M^\prime p$, where $M^\prime$ is the same as matrix $M$ in \eqref{intmat} except that $diag(M^\prime) = [1,1,...,1]^T$. It should be noted that in problem $VI(\mathcal{Q}, F)$, whenever the function $F(x)$ is an affine function, there is no difference between strictly monotonicity and strongly monotonicity \cite{pang}.
Using theorem 3 we deduce that:
\newtheorem{corollary}{\bf{Corollary}}
\begin{corollary}
The problem $VI(\mathcal{Q},F)$, with $F(p) = M^\prime p$, has a unique solution, if the same uniqueness conditions derived in theorem 1 can be satisfied.
\begin{IEEEproof}
Let $D(d_i, [M^\prime]_{ii})$ be the closed disc centered at $[M^\prime]_{ii}$ with radius $d_i = \sum_{j\neq i}\left|[M^\prime]_{ij}\right|$.  Using the Gershgorin circle theorem, 
$\forall i \le max\left\{(\sum_{q=1}^Q\nu_q) , (\sum_{q=1}^QN_{T_q})\right\}$, every eigenvalue of $M^\prime$ is within at least one of the aforementioned circles. We also know that for the function $F$, in order to be strictly (or strongly) monotone, the matrix $M^\prime$ has to positive definite. Hence, all the radii of the Gershgorin circles must be less than one to ensure that the matrix $M^\prime$ remains positive definite. Using this fact (for $M^\prime$ and ${M^\prime}^T$), the conditions written in \eqref{th1} and \eqref{th2} can be easily derived.
\end{IEEEproof}
\end{corollary}
{\bf Remark:} From the variational inequality point of view, we can again show that the conditions in \eqref{th1} and \eqref{th2} are not tight enough to predict the uniqueness of NE. This can be shown by the fact that if the radii of Gershgorin circles become larger than one, the probability of having positive eigenvalues (i.e., positive definiteness of $M^\prime$) still exists. However, the derived conditions in \eqref{th1} and \eqref{th2} skip this probability in order to introduce more practical conditions.

The power control game introduced as a variational inequality, can help us design modified algorithms that improve the performance of the network. The next section deals with designing such modified algorithms.
\section{Modifying Algorithms}
After introducing the alternative definition of power control game, we present our modified algorithms in this section in order to overcome the issues related to our power control method.
\subsection{Regularization: Guaranteeing Convergence}
\label{regu}
The physical interpretation of the previous uniqueness conditions implies that sum of normalized received (generated) interference at (by) each receive (transmit) antenna must be lower than one. In the cases where such condition is not satisfied, there will be more than one NE, and the proposed algorithm is not guaranteed to converge. In order to overcome this issue, we add the term $\frac{\tau}{2} \left|\left|{p^i_q}^{(n)}-{p^i_q}^{(n-1)}\right|\right|^2$ to the objective function of \eqref{optequi}, while we are at the $n$th iteration of finding the NE ($\tau > 0$). By adding this term, and after writing KKT conditions again, we have the problem $VI(\mathcal{Q}, F_\tau)$, where $F_\tau(p) = (M^\prime +\tau I)p+b$, and $b = -\tau p^{(n-1)} $. In fact, by adding the (large enough) constant $\tau$ to the diagonal elements of the matrix $M^\prime$, we move the centers of Gershgorin circles farther from the origin (to larger positive values) so that the radii of the circles can be larger than one, and the positive definiteness of the Jacobian matrix of $F_{\tau}$, namely $\mathcal{J}_{\tau}$, can be guaranteed more than before. After the modification of objective function, the waterfilling solution changes to 
\begin{equation}
\label{water2}
{p^i_q}^* = \left(\frac{\mu_q}{\tau+1}-\frac{c^i_q}{\tau+1}+\frac{\tau}{\tau+1}{p^i_q}^{(n-1)}\right)^+.
\end{equation}
Therefore, the third step of algorithm \ref{alg} should be modified accordingly. The free parameter $\tau$ must satisfy
\begin{equation}
\tau>\max_{(i,q)}\sum_{r\neq q}\sum_{j=1}^{N_{T_r}} \frac{\left|\left[\tilde{H}_{rq}\right]_{i,j}\right|^2}{|\sigma_q^i|^2}-1,
\end{equation}
This constraint ensures the uniqueness of NE along with the convergence of the asynchronous algorithm, and can be easily calculated by finding the limit (or the distance) below which the matrix $M^\prime+\tau I$ is positive definite. Depending on how much coordination exists between the users, the constant $\tau$ can be estimated\footnote{Adaptive estimation of $\tau$ can be a potential subject of future work.}. It should be noted that choosing $\tau$ to be too much large can reduce the effect of interference in adjusting the power allocation and increases the effect of previous iterations in power allocation, which is not of our interest.
It is important to know that that whether the solution found for $VI(\mathcal{Q}, F_\tau)$ is in the solution set of $VI(\mathcal{Q}, F)$. If a vector $p^*$ is a solution for $VI(\mathcal{Q}, F_\tau)$, for all $p\in\mathcal{Q}$ we have
\begin{equation}
(p-p^*)^TF_\tau(p^*)\ge0.
\end{equation}
Because we ensured that by a proper choice of $\tau$, the convergence to the solution of $VI(\mathcal{Q}, F_\tau)$ is guaranteed, which means ${p^i_q}^{(n)} \simeq {p^i_q}^{(n-1)}$ when $n \rightarrow \infty$. Hence, at the convergence point, for all $p\in\mathcal{Q}$ we have:
\begin{align}
(p-p^*)^T\left(F(p^*)+\tau(p^*-p^*)\right)\ge0,
\notag
\\
\Rightarrow~(p-p^*)^TF\left(p^*\right)\ge0.
\end{align}
Therefore, the solution found for $VI(\mathcal{Q}, F_\tau)$ is also a solution for $VI(\mathcal{Q},F)$.
\subsection{Controlling the Convergence to the NE}
\label{meri}
Despite guaranteeing the convergence of the power control game in the previous method, there is still no control on which NE the game converges to. In other words, there is still no guarantee on the efficiency of the achieved NE. The solution to this problem is that we have to choose the most efficient NE in our power control game. Meaning that the chosen NE should optimize a particular criterion. Considering the solution set of $VI(\mathcal{Q},F)$ as $SOL(\mathcal{Q},F)$, we want to reach to the NE point that optimizes a convex function $\phi(p)$. To do so, we have the following optimization:
\begin{align}
\label{merit}
\bf{\min_{p}}~~\phi(p)~~~~~~~~~~~
\notag
\\
\textrm{subject to }~p\in \text{SOL}(\mathcal{Q}, F).
\end{align}
As there is no prior access to the $SOL(\mathcal{Q},F)$ (i.e., no access to all of the equilibrium points), this optimization cannot be done using bi-level optimization algorithms. We propose to solve the problem $VI(\mathcal{Q},F_{\epsilon,\tau})$, where:
\begin{equation}
\label{control}
F_{\epsilon,\tau}(p) = F(p)+\epsilon^{(n)}\nabla\phi(p)+\tau({p}-{p}^{(n-1)}),
\end{equation}
and $\nabla\phi(p) = \left[\nabla_{p_1}\phi(p),...,\nabla_{p_Q}\phi(p)\right]^T$ is the gradient vector of the function $\phi(p)$ with respect to the vector $p$, $\epsilon^{(n)}$ is a positive variable that is a function of iteration number, and the term $\tau({p}-{p}^{(n-1)})$ is in fact the same modification done in the previous subsection. It is clear that if the game reaches to the NE that minimizes $\phi(p)$, the second term of \eqref{control} has zero (or the smallest) value. The most important contribution of \eqref{control} is summarized in the following theorem:
\begin{theorem}
Consider the monotone problem $VI(\mathcal{Q},F)$ (i.e., $F(p)$ is monotone in this problem) with the solution set $SOL(\mathcal{Q},F)$. For $n \rightarrow \infty$, the solution of the problem $VI(\mathcal{Q},F_{\epsilon, \tau})$ converges to the minimum point of $\phi(p)$, if the following assumptions hold:
\begin{enumerate}
\item{}
The solution set $SOL(\mathcal{Q},F)$ is closed and convex.
\item{}
$\phi(p)$ is Lipschitz continuous on $SOL(\mathcal{Q},F)$ with the modulus $L_\phi$.
\item{}
the constant $\tau$ satisfies the following constraint
\begin{equation}
\tau>\max_{(i,q)}\sum_{r\neq q}\sum_{j=1}^{N_{T_r}} \frac{\left|\left[\tilde{H}_{rq}\right]_{i,j}\right|^2}{|\sigma_q^i|^2}+\epsilon^{(n)}L_\phi-1
\end{equation}
\item{}
The constant $\epsilon^{(n)}$ is chosen such that $lim_{n\rightarrow \infty} \epsilon^{(n)} = 0$, and $\sum_{n=1}^\infty \epsilon^{(n)} = \infty$.
\end{enumerate}
\begin{IEEEproof}
See Appendix B.
\end{IEEEproof}
\end{theorem}
Considering the above theorem, we can design a framework in which the users can optimize a certain merit function, while non-cooperatively competing with each other. The following algorithm describes how this framework can be implemented:
\begin{algorithm}[H]
\caption{}
\label{algcontrol}
Choose suitable 
$\epsilon^{(n)}$ and $\tau$ and any $p^{(0)} \in \mathcal{Q}$
\begin{algorithmic}[1]
\For{n=1 to $outerit_{max}$}
\Repeat
\State
$p^{(n)}$ = Solve $VI(\mathcal{Q},F_{\epsilon,\tau})$~
\texttt{//Use Jacobian}
\Until{Convergence}
\EndFor
\end{algorithmic}
\end{algorithm}
It is clear that this algorithm is comprised of two loops of iterations. The third line is the inner loop of iterations, in which users non-cooperatively optimize their objective functions. In other words, the third line is the same as the non-cooperative game implemented in algorithm \ref{alg} with the difference that  the waterfilling solution changes to \begin{equation}
\label{water3}
{p^i_q}^* = \frac{1}{\tau+1}\left(\mu_q-c^i_q+\tau{p^i_q}^{(n-1)}-\epsilon^{(n)}\nabla^i_{p_q}\phi(p^{(n)})\right)^+.
\end{equation}
Therefore, the third step of algorithm \ref{alg} should be modified accordingly. Using the third assumption in theorem 4, in each iteration of the above algorithm, we can use Jacobian method for the third line\footnote{Note that due to the practical limitations imposed on the network, the asynchronous algorithm introduced in the previous section can be used.}.
\subsection{Inexact Convergence}
Although the theory behind previous modifications might seem to increase the performance of non-cooperative power control games, running the algorithm VII.1 requires a lot of iterations which makes this algorithm too slow to be implemented. In this part of paper, we propose a solution in which we calculate an inexact solution of the $VI(\mathcal{Q},F_{\epsilon, \tau})$, allowing us to run algorithm \label{algcontrol} with a lower number of iterations. 
The following property has an important contribution in deducing the inexact convergence.
\begin{proposition}
Consider the monotone problem $VI(\mathcal{Q},F_\tau)$, that has solution set $SOL(\mathcal{Q},F_\tau)$. Given the inexact solution set $SOL_{\delta_n}(\mathcal{Q},F_\tau)$ in the $n$th iteration as
\\
\\
\begin{align}
\label{soldelta}
SOL_{\delta_n}(\mathcal{Q}, F_\tau) = \left\{ p^{(n)}_\delta \in\mathcal{Q}\left| F_\tau(p_\delta^{(n)})^T(y-p_\delta^{(n)})\ge -{\delta_n}\right.\right\}
\notag
\\
\forall y\in \mathcal{Q},
\end{align}
\\
and assuming $\lim_{n\rightarrow\infty}\frac{\delta_n}{\tau} = 0$ and $\sum_{n=0}^{\infty}\frac{\delta_n}{\tau} < \infty$, we have:
\begin{equation}
\lim_{n\rightarrow \infty}||p_\delta^{(n)}-p_\delta^{(n-1)}|| = 0.
\end{equation}
\begin{IEEEproof}
See Appendix C.
\end{IEEEproof}
\end{proposition}
As the choice of $F$ (or $F_\tau$) in the above proprty was arbitrary, we can draw the same conclusion for $F_{\epsilon}(p) = F(p)+\epsilon^{(n)}\nabla\phi(p)$ (or $F_{\epsilon, \tau}$) as well if the term $\nabla\phi(p)$ is also monotone\footnote{In the next subsection, the choices for the merit function satisfies monotonicity. The proofs for the monotinicity of the functions used in simulation part are skipped for brevity.}. The result of proposition 3 makes us ready to present the main theorem regarding inexact convergence of power control game.
\begin{theorem}
Consider the monotone problem $VI(\mathcal{Q},F_{\epsilon, \tau})$, that has solution set $SOL(\mathcal{Q},F_{\epsilon, \tau})$. Given the inexact solution set $SOL_{\delta_n}(\mathcal{Q},F_{\epsilon, \tau})$ in the $n$th iteration as:
\\
\\
\begin{align}
SOL_{\delta_n}(\mathcal{Q}, F_{\epsilon, \tau}) = \left\{ p^{(n)}_\delta \in\mathcal{Q}\left| F_{\epsilon, \tau}(p_\delta^{(n)})^T(y-p_\delta^{(n)})\ge -{\delta_n}\right.\right\}
\notag
\\
\forall y\in \mathcal{Q},
\end{align}
\\
and assuming $\lim_{n\rightarrow\infty}\frac{\delta_n}{\tau} = 0$ and $\sum_{n=0}^{\infty}\frac{\delta_n}{\tau} < \infty$, for $n \rightarrow \infty$, the sequence of the solutions $p_\delta^{(n)}\in SOL_{\delta_n}(\mathcal{Q},F_{\epsilon, \tau})$ converge to the minimum point of $\phi(p)$
\begin{IEEEproof}
See Appendix D.
\end{IEEEproof}
\end{theorem}
The result of the analyses done in this subsection, provides us with the following algorithm:
\begin{algorithm}[H]
\caption{Inexact Power Control Algorithm}
\label{alginexact}
Choose suitable 
$\epsilon^{(n)}$ and $\tau$ and any $p^{(0)} \in \mathcal{Q}$
\begin{algorithmic}[1]
\For{n=1 to $outerit_{max}$}
\Repeat
\State
Find $p_\delta^{(n)}\in SOL_{\delta_n}(\mathcal{Q},F_{\epsilon,\tau})$, that satisfies $F_{\epsilon, \tau}(p_\delta^{(n)})^T(y-p_\delta^{(n)})\ge -{\delta_n},~\forall y\in \mathcal{Q}$
\\
~~~~~~~~~~~~~~~~~~~~~~~~~~~~~~~~~~~~~~~~~~~~~~~\texttt{//Use Jacobian}
\Until{Convergence}
\EndFor
\end{algorithmic}
\end{algorithm}
\begin{figure}
\centerline{
\includegraphics
[scale = 0.25, trim = 25mm 10mm 3mm 29mm]
{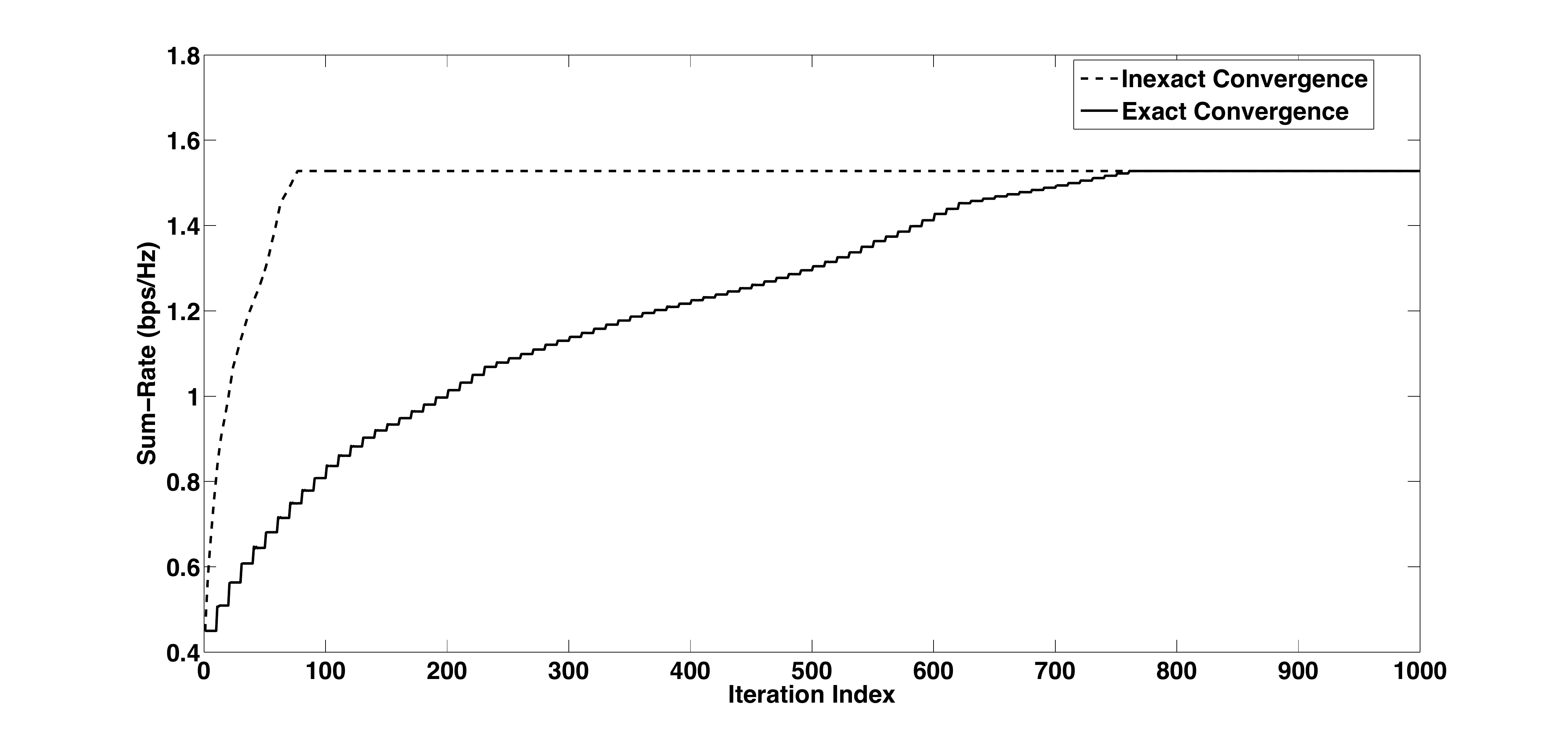}
}
\caption{\small Comparison of convergence speed of the modified algorithms}
\label{ive}
\end{figure}
\begin{figure}
\centerline{
\includegraphics
[scale = 0.25, 
trim = 30mm 10mm 10mm 10mm
]
{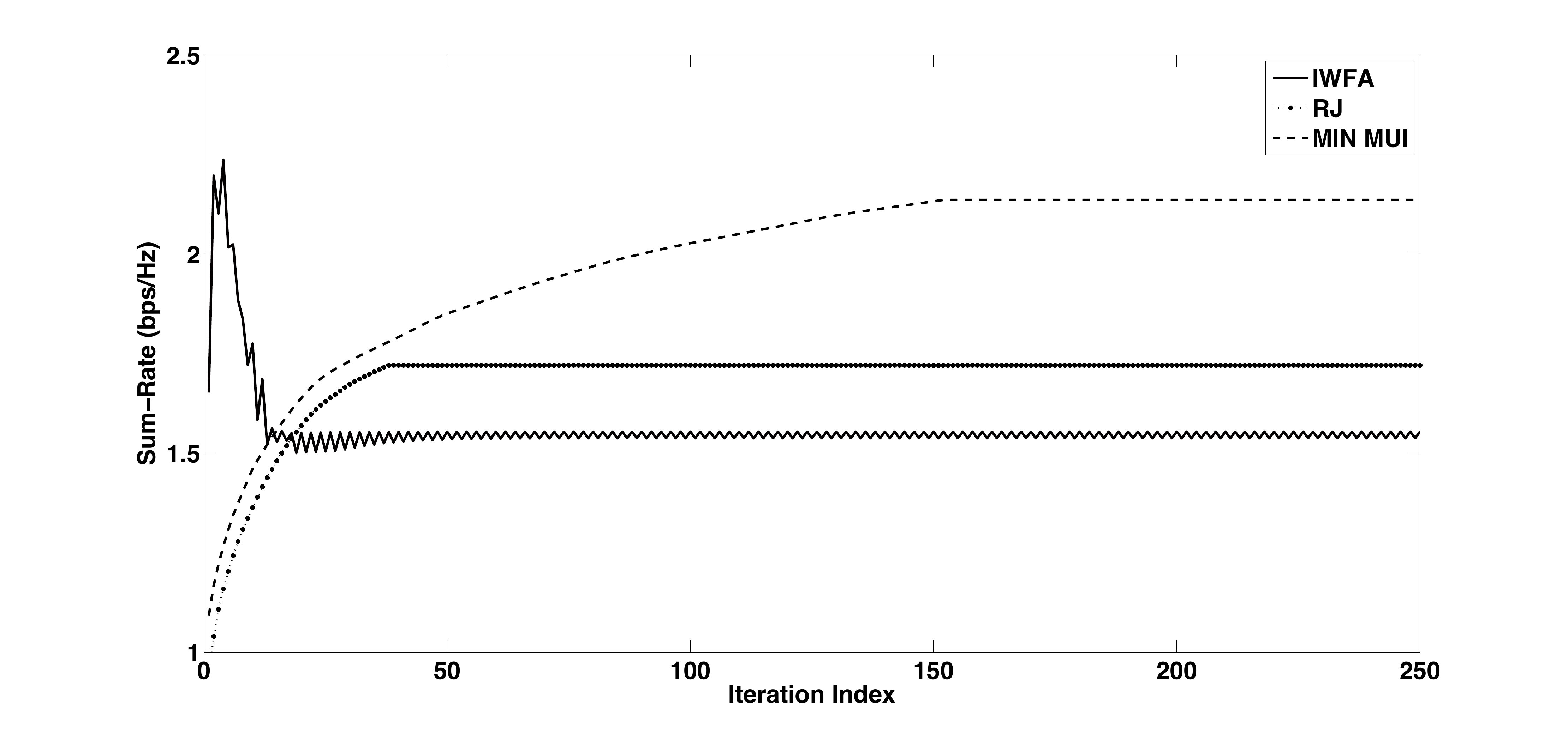}
}
\caption{\small Comparison of convergence and the performance of the proposed algorithms}
\label{improve}
\end{figure}
\subsection{On the Choice of Merit Function}
Choosing the merit function is the important part of the modification, as it determines the amount of signaling, computation, and performance improvement. We mention two options for a possible merit function. These two merit functions have different performance from one another and impose different signaling requirement.
\subsubsection{Minimizing Multiuser Interference}
The sum of the multiuser interference is the sum of the rows of the interference function $I(p) = M.p$ (see Appendix A.). Therefore, by using the merit function as
\begin{equation}
\label{muimerit}
\phi(p) = \sum_{q = 1}^Q\sum_{i}\sum_{r\neq q}\sum_{j = 1}^{N_{T_q}}\left|\left[\tilde{H}_{rq}\right]_{i,j}\right|^2p^j_r,
\end{equation}
the power control game ends up with the NE that minimizes total multiuser interference. It should be noted that the index $i$ counts the rows of $M$ in \eqref{intmat}. Hence, the elements of $\nabla_{p_q}\phi(p)$ --namely as $\nabla^i_{p_q}\phi(p)$-- can be shown as
\begin{equation}
\nabla^i_{p_q}\phi(p) = \sum_j\sum_{r \neq q}\left|\frac{\left[\tilde{H}_{qr}\right]_{j,i}}{\sigma^j_r}\right|^2,~i = 1,..., N_{T_q},
\end{equation}
where $j$ counts the rows of matrix $M$ shown in \eqref{intmat}. Equation \eqref{muimerit} can be used in the waterfilling operator (i.e.,\eqref{water2}) accordingly.
\subsubsection{Maximizing Sum-rate}
Let the merit function for
\\$i = 1,...,N_{T_q}$ be expressed as
\begin{align}
\label{prices}
\nabla_{p_q}^i\phi(p_q) = \sum_j\sum_{r\neq q}\left|\frac{\left[\tilde{H}_{qr}\right]_{j,i}}{\sigma^j_r}\right|^2\frac{p^{j}_r}{c^{j}_r(~c^{j}_r+p^{j}_r~)},
\end{align}
where $j$ counts the row of of the matrix $M$ shown in \eqref{intmat}. 
using \eqref{prices} in \eqref{water2}, we end up with the same problem that the previously proposed pricing algorithms were trying to solve \cite{yu1, yu2, ghamari}. The difference in here is that we determined the condition to implement the pricing algorithm asynchronously, while the previous studies proposed Jacobi and Gauss-Seidel methods without any proof of convergence for asynchronous method. The merit function that the users try to optimize is
\begin{equation}
\phi(p) = \left[-\sum_{r\neq 1}R_r, -\sum_{r\neq 2}R_r, ..., -\sum_{r\neq Q}R_r \right]^T.
\end{equation}
Therefore, the intuitive meaning of the pricing algorithms is that while a user optimizes its power allocation, it tries to use a power allocation that maximizes sum-rate of other users.
\section{Simulation Results and Discussion}
In this section, we verify our theoretical analyses done in the previous section. We assumed the same setting as in section VI for the MIMO channel plus the assumption that the MIMO channel is frequency selective (with a channel order of $L = 3$), so we set 16 subcarriers for each user. There are 10 users, each with a 2x2 direct MIMO channel.
\subsection{Exact vs. Inexact Convergence}
Fig. \ref{ive} shows the sum-rate of the users versus the number of iterations for exact and inexact approach in one channel realization. The sequence of errors $\delta_n$ is set to $\delta_n = (0.95)^n$. We used Jacobi method for inner loop of algorithms VII.1 and VII.2 whenever they are simulated. The minimization of multiuser interference (with $\epsilon^{(n)} = \frac{1}{1+10n}$) is used as the merit function. It is clear that using inexact convergence can considerably reduce the number of iterations.
\subsection{Improving the sum-rate of the users}
Fig. \ref{improve} shows the sum-rate of the users versus the number of iterations for 3 different algorithms in one channel realization. The solid curve (labeled "IWFA") shows the simple asynchronous algorithm described in algorithm \ref{alg}, the dotted curve (labeled "RJ") shows the modified algorithm designed in section \ref{regu}, and the dashed curve (labeled "MIN MUI") shows a controlled convergence designed in section VII-B that uses minimum multiuser interference as its merit function. 
\begin{figure}
\centerline{
\includegraphics
[scale = 0.27, 
trim = 35mm 10mm 3mm 29mm
]
{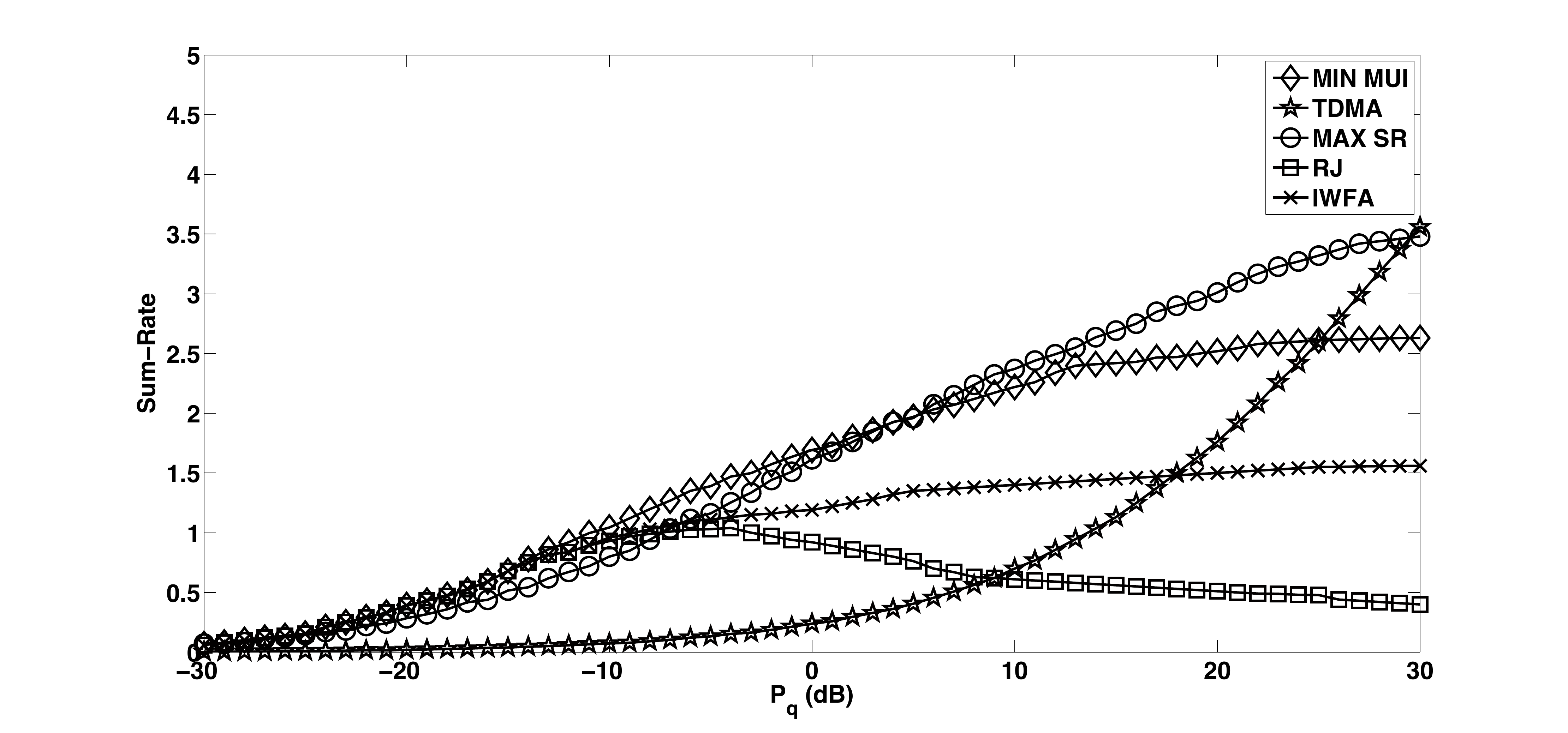}
}
\caption{\small Comparison of sum-rate of the proposed algorithms for different transmit powers}
\label{sum1}
\end{figure}
\begin{figure}
\centerline{
\includegraphics
[scale = 0.27, 
trim = 35mm 10mm 3mm 20mm
]
{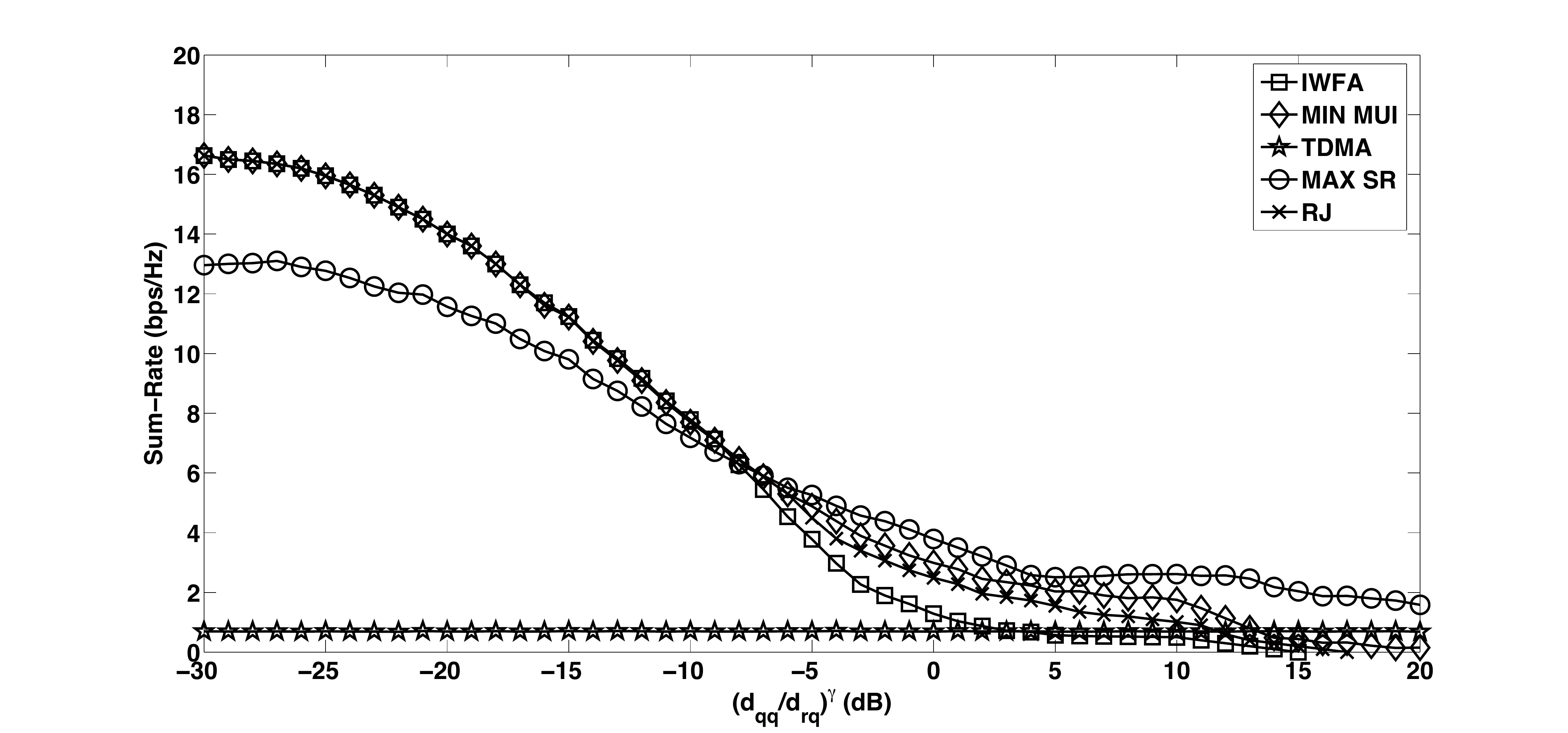}
}
\caption{\small Comparison of sum-rate of the proposed algorithms for different values of path-loss}
\label{sum2}
\end{figure}
It can be seen that IWFA method oscillates between two points. In such condition the interference is strong and the conditions in theorem 1 are not satisfied, so there are multiple equilibria in the power control game. The multiple equilibria appears as oscillation in the convergence trend of IWFA algorithm. It is clear that both modified algorithms converge eventually. Furthermore, we see that (despite a few signaling between the users) controlling convergence improves the performance more than when we do not have a merit function. Thus, because of the absence of a merit function in RJ algorithm, we may not necessarily improve the performance of the network.
\subsection{Comparison of Sum-rate}
In Fig. \ref{sum1}, the sum-rate of the users (in bps/Hz) versus the total transmit power is depicted. It is assumed that all the transmitters have the same total transmit power. The path-loss for each user is fixed at $\left(\frac{d_{qq}}{d_{rq}}\right)^\gamma = 10 dB$, and $P_q^i = 10 dB$ (see proposition 2). Each point on the curves is the mean value of 1000 channel realizations. The curve labeled "MAX SR" shows the modified algorithm with maximization of the sum-rate as the merit function, and the curve labeled "TDMA" shows the performance of the network when the interference free approach --namely as "Time Division Multiple Access"-- is used.
We see that the MAX SR algorithm has a performance loss for low transmit power and the highest performance for larger transmit powers. The loss of performance occurs because of the fact that the elements of $\nabla \phi(p)$ are large for low transmit powers which makes users' power allocation dominated by the merit function. In fact, users do not use their transmit power efficiently in channels as the price they pay is too much expensive. However, this trend vanishes and the advantage of using MAX SR emerges as the total transmit power of the users grow. MIN MUI algorithm performs better than IWFA in all of the transmit powers. IWFA algorithm has a better performance than the TDMA method, but its advantage is lost as the transmit power grows, because the selfish act of the users grows interference in the network. RJ algorithm performs better than IWFA, but because of no control on its convergence point, it cannot reach the performance of MAX SR and MIN MUI.

Fig. \ref{sum2} shows the sum-rate of the users versus the path-loss (i.e., $\left(\frac{d_{qq}}{d_{rq}}\right)^\gamma$), when the total transmit power is fixed at $P_q = 10 dB$. It can be seen that for low interference powers ($\left(\frac{d_{qq}}{d_{rq}}\right)^\gamma$ is small) MAX SR has some performance degradation. Because in low interference power, the users are almost orthogonal to each other, this interference management technique is not very much superior to simpler techniques. However, for higher interference powers, the performance of MAX SR becomes more evident.
\section{Conclusion}
In this paper, the concept of contractive functions was employed to derive the convergence conditions of the distributed power control in multi-user MIMO networks. The convergence conditions derived in this paper can be used in practice by the designer to create a distributed power control framework. Simulations show that the proposed criteria can predict a unique NE more than the previous work. Furthermore, the proposed power control algorithm can be implemented asynchronously, which gives a noticeable flexibility to our algorithm depending on the practical limitations. Despite these advantages, we modified our algorithm to improve the sum-rate of the network. It was shown that by exchanging a limited amount of signaling, improving the performance of our algorithm is possible. Furthermore, we proposed the inexact version of our modified algorithm to boost the convergence speed. Although the inexact approach seems promising in reducing convergence time, lowering the amount coordination needed for the modified approaches must be taken into account as an interesting subject of future research.
\appendices
\section{Proof of Theorem 1}
\begin{figure*}[!t]
\small
\begin{equation}
\label{intmat}
\hspace{-7mm}
M = 
\begin{matrix}
 \begin{bmatrix}
\bovermat{$N_{T_1}$}{&~~~~~~~~~&~~~~~~~~~~~~~~~~~&}
\\[-2.3em]
0 & \cdots & 0 &
\left|\frac{\left[\tilde{H}_{21}\right]_{1,1}}{\sigma^1_1}\right|^2 & \cdots & \left|\frac{\left[\tilde{H}_{21}\right]_{1,N_{T_2}}}{\sigma^1_1}\right|^2 & \left|\frac{\left[\tilde{H}_{31}\right]_{1,1}}{\sigma^1_1}\right|^2 & \cdots & \left|\frac{\left[\tilde{H}_{Q1}\right]_{1,N_{T_Q}}}{\sigma^1_1}\right|^2 
\\[1.5em]
0 & ~~~~\cdots~~~~ & 0 & \left|\frac{\left[\tilde{H}_{21}\right]_{2,1}}{\sigma^2_1}\right|^2 & \cdots & \left|\frac{\left[\tilde{H}_{21}\right]_{2,N_{T_2}}}{\sigma^2_1}\right|^2 & \left|\frac{\left[\tilde{H}_{31}\right]_{2,1}}{\sigma^2_1}\right|^2  & \cdots & \left|\frac{\left[\tilde{H}_{Q1}\right]_{2,N_{T_Q}}}{\sigma^2_1}\right|^2 
\\[0.8em]
\vdots&&\vdots&\vdots&&\vdots&\vdots&&\vdots
\\[0.8em]
0 & ~~~~\cdots~~~~ & 0& \left|\frac{\left[\tilde{H}_{21}\right]_{\nu_1,1}}{\sigma^{\nu_1}_1}\right|^2 & \cdots & \left|\frac{\left[\tilde{H}_{21}\right]_{\nu_1,N_{T_2}}}{\sigma^{\nu_1}_1}\right|^2 & \left|\frac{\left[\tilde{H}_{31}\right]_{\nu_1,1}}{\sigma^{\nu_1}_1}\right|^2  & \cdots & \left|\frac{\left[\tilde{H}_{Q1}\right]_{\nu_1,N_{T_Q}}}{\sigma^{\nu_1}_1}\right|^2 
\\[3em]
&&&\bovermat{$N_{T_2}$}{&~~~~~~~&~~~~~~~~~~~~~~~~~}
\\[-3.5em]
\\
\left|\frac{\left[\tilde{H}_{12}\right]_{1,1}}{\sigma^{1}_2}\right|^2 & \cdots & \left|\frac{\left[\tilde{H}_{12}\right]_{1,N_{T_1}}}{\sigma^{1}_2}\right|^2
& 0 & \cdots & 0 & \cdots & \cdots & \left|\frac{\left[\tilde{H}_{Q2}\right]_{1,N_{T_Q}}}{\sigma^{1}_2}\right|^2
\\[0.8em]
\vdots&&\vdots&\vdots&&\vdots&\vdots&&\vdots
\\[0.8em]
\left|\frac{\left[\tilde{H}_{12}\right]_{\nu_2,1}}{\sigma^{\nu_2}_2}\right|^2 & \cdots & \left|\frac{\left[\tilde{H}_{12}\right]_{\nu_2,N_{T_1}}}{\sigma^{\nu_2}_2}\right|^2
& 0 & \cdots & 0 & \cdots & \cdots & \left|\frac{\left[\tilde{H}_{Q2}\right]_{\nu_2,N_{T_Q}}}{\sigma^{\nu_2}_2}\right|^2
\\[0.8em]
\vdots&&\vdots&\vdots&&\vdots&\vdots&&\vdots
\\[0.8em]
&&&&&&\bovermat{$N_{T_Q}$}{&~~~&~~~~~~~~~~~~~~~~~}
\\[-3.5em]
\\
\left|\frac{\left[\tilde{H}_{1Q}\right]_{1,1}}{\sigma^{1}_Q}\right|^2 & \cdots & \left|\frac{\left[\tilde{H}_{1Q}\right]_{1,N_{T_1}}}{\sigma^{1}_Q}\right|^2
&
\left|\frac{\left[\tilde{H}_{2Q}\right]_{1,1}}{\sigma^{1}_Q}\right|^2 &  \cdots & \cdots
& 0 & \cdots & 0
\\[0.8em]
\vdots&&\vdots&\vdots&&\vdots&\vdots&&\vdots
\\[0.8em]
\left|\frac{\left[\tilde{H}_{1Q}\right]_{\nu_Q,1}}{\sigma^{\nu_Q}_Q}\right|^2 & \cdots & \left|\frac{\left[\tilde{H}_{1Q}\right]_{\nu_Q,N_{T_1}}}{\sigma^{\nu_Q}_Q}\right|^2
&
\left|\frac{\left[\tilde{H}_{2Q}\right]_{\nu_Q,1}}{\sigma^{\nu_Q}_Q}\right|^2 & \cdots  & \cdots
& 0 & \cdots & 0
  \end{bmatrix}
 \begin{aligned}
\\[-3em]
\\
\left.
\begin{matrix}
\\ \\ \\ \\ \\ \\ \\ \\ \\ \\
  \end{matrix} \right\} \nu_1
 \\ \\
\left.
\begin{matrix}
\\ \\ \\ \\ \\ \\ \\
  \end{matrix} \right\}\nu_2
\\
\\[-1em]
\vdots~~~~
\\
\\[-.6em]
\left.
\begin{matrix}
\\ \\ \\ \\ \\ \\
  \end{matrix} \right\} \nu_Q
 \end{aligned}
\\
\\
\end{matrix}
\end{equation}
\hrulefill
\end{figure*}
Assuming $p = [p_1, p_2, ..., p_q, ..., p_Q]^T$ with $p_q = [p_q^1, ...,p_q^{N_{T_q}} ]^T = diag\left(\mathcal{E}\left\{ x_qx_q^H\right\}\right)$, we first prove that the interference function $I(p) = [I_1(p), ..., I_Q(p)]^T$ is contractive, where $I_q(p) = [I^1_q(p),...,I^{\nu_q}_q(p)]^T$ and $I^i_q(p) =c^i_q$. Next, we use the second item in proposition 1 to prove that the water-filling operator is also a contractive function\footnote{The water-filling operator can be achieved by a simple manipulation of $I^i_q(p)$. Without the loss of generality of our approach, we negelected the second term in the representation of $c^i_q$ during the following analyses.}.

The contractivity of the interference functions can be shown if we are able to show the vector $I(p)$ with a close-form representation.The interference function $I(p)$ can be shown as
\begin{equation}
I(p) = M.p,
\end{equation}
where $M$ is a $\{(\sum_{q=1}^Q\nu_q)\} \times \{(\sum_{q=1}^QN_{T_q})\}$ matrix which is shown in \eqref{intmat}.
Depending on the value of $\nu_q$, we present different proofs for contractivity.

{\bf 1) $\bf \nu_q = N_{T_q} = N_{R_q}$:}
In this case, the matrix $M$ is a square matrix. While checking the contractivity properties of $I(p)$, we conclude that the first two properties (i.e. positivity and monotonicity) are obvious. For the third property (i.e. contractivity property), assuming a positive vector $v$, we have
\begin{equation}
\label{cont}
I(p+\epsilon v) = M.p + M.\epsilon v \le I(p) + ||M||_\infty^v \epsilon v,
\end{equation}
where $||.||_\infty^v$ is the weighted maximum norm given the vector $v > 0$. Therefore, if $||M||_\infty^v < 1$, the function $I(p)$ is a contractive function, and according to the properties of contractive functions, the iterative water-filling between the users will converge to a unique fixed point that is the Nash equilibrium of the power control game. It is known that for the non-negative square matrices (e.g. $M$), there exists a positive vector $v$ such that $||M||_\infty^v < 1$ if and only if $\rho(M) <1$, where $\rho (.)$ is the spectral radius of a matrix \cite{feyz13}. Both of these measures are difficult to evaluate in the practical cases because the designer has to have access to the matrix $M$ to predict the uniqueness of NE. The easiest verifiable conditions can be derived by choosing $v = 1$, which yields the inequality in \eqref{th1}, or equivalently $||M||_\infty < 1$. Since $\rho(M) = \rho(M^T)$, all of the aforementioned derivations about  a non-negative square matrix can be done for its transposed version (i.e. $M^T$). Doing so for $M^T$ yields the inequality in \eqref{th2}. The physical interpretation of \eqref{th1} and \eqref{th2} suggests that the sum of normalized interference produced by each user has to be less than one, or alternatively the sum of received normalized interference at each receiver should be less than one. This intuitive interpretation can be easily used by the designer to determine a minimum distance that ensures the uniqueness of NE with high probability.

{\bf 2) $\bf \nu_q = N_{T_q} < N_{R_q}$:} As \eqref{intmat} suggests, in this case, the matrix $M$ is still a square matrix and all of the aforementioned proof about the previous case is also true about this case.

{\bf 3) $\bf \nu_q = N_{T_q} > N_{R_q}$:} In this case, the matrix $M$ is not square and the proofs of the previous cases cannot be employed for this case. As we want to have a set of unified conditions for all the cases, a simple manipulation of the matrix $M$ can make it a square matrix for this case. This manipulation is done by adding several rows of zeroes to the matrix $M$ for the user --say $q$th user-- that has $N_{T_q} > N_{R_q}$. Adding zero rows is done until sum of the $q$th user's transmit antennas equals to the sum of both its receive antennas and the zero rows added for the $q$th user. We should note that by adding rows of zeroes for each user (whose transmit antennas are more than its receive antennas), the whole measurements for the interference will not be affected. In fact, we assume that adding zero rows to $M$ is equal to adding a receive antenna that does not receive neither signal nor interference (i.e. it is turned off). Hence, the matrix $M$ is eventually square again and we can use the same proofs we used previously.
%
\section{Proof of Theorem 4}
First of all, we explain the reason behind making the assumptions in this theorem. The first assumption is made so that the optimization problem in \eqref{merit} will be a convex optimization. The second assumption puts an upper limit on the variations of $\phi(p)$ so that we can calculate the constant $\tau$ and make the third assumption to ensure the uniqueness of NE\footnote{One can check that all of the merit functions chosen in this paper satisfy Lipschitz continuity.}. It may seem that the Lipschitz modulus cannot be calculated in practical situations. Because of this reason, the constant $\epsilon^{(n)}$ is chosen as a terminating sequence. Hence, as the iterations grow up, the term $\epsilon^{(n)}\nabla^i_{p_q}\phi(p)$ vanishes. However, this term should not be terminated too much fast because otherwise, the effect of this term (minimizing $\phi(p)$) would not be significant. Therefore, we make the assumption $\sum_{n=1}^\infty\epsilon^{(n)} = \infty$ to have a slowly terminating sequence. Examples of such sequence are:
\begin{itemize}
\item{}
$\epsilon^{(n)} = \epsilon^{(n-1)}\left(1-\mathcal{E}\epsilon^{(n-1)}\right),~~\mathcal{E}\in(0,1)$.
\item{}
$\epsilon^{(n)} = \frac{1}{1+\alpha n}
, ~~ \alpha >0$.
\end{itemize}

We let the solution set of \eqref{merit} as $\mathcal{S} \subset SOL(\mathcal{Q},F)$. While in the $n$th iteration, let us introduce the mapping $G_{\mathcal{S}}(p^{(n-1)})$ as the euclidean projection of the vector $p^{(n-1)}$ on the set $\mathcal{S}$. As $G_\mathcal{S}(p^{(n-1)})\in \mathcal{S} \subset SOL(\mathcal{Q},F)$, by considering $VI(\mathcal{Q},F)$in the $n$th iteration we have:
\begin{equation}
F(G_{\mathcal{S}}(p^{(n-1)}))^T(p^{(n)}-G_{\mathcal{S}}(p^{(n-1)}))\ge 0.
\end{equation}
We know that $F(p)$ is monotone, so:
\begin{equation}
\label{mono}
F(p^{(n)})^T(G_{\mathcal{S}}(p^{(n-1)})-p^{(n)})\le 0.
\end{equation}
Replacing $p^{(n)}$ in $VI(\mathcal{Q},F_{\epsilon, \tau})$ we have:
\begin{align}
\left[ F(p^{(n)})+\epsilon^{(n)}\nabla\phi(p^{(n)})\right]^T(y-p^{(n)})\ge 
\notag
\\
\tau(p^{(n-1)}-p^{(n)})(y-p^{(n)}).
\end{align}
If the third assumption holds, $\lim_{n\rightarrow \infty}||p^{(n)}-p^{(n-1)}|| = 0$. Hence, 
\begin{equation}
\left[ F(p^{(n)})+\epsilon^{(n)}\nabla\phi(p^{(n)})\right]^T(y-p^{(n)})\ge 0.
\end{equation}
setting $y = G_\mathcal{S}(p)$ and considering \eqref{mono}:
\begin{align}
\label{expand1}
0 \ge F(p^{(n)})^T(G_{\mathcal{S}}(p^{(n-1)})-p^{(n)})\ge 
\notag
\\
\epsilon^{(n)}\nabla\phi(p^{(n)})^T(p^{(n)}-G_{\mathcal{S}}(p
^{(n-1)})),
\notag
\\
\notag
\\
\Rightarrow~~\nabla\phi(p^{(n)})^T(G_{\mathcal{S}}(p
^{(n-1)})-p^{(n)})\ge 0.
\end{align}
Therefore, as $n \rightarrow \infty$, the inequality in \eqref{expand1} shows the minimum principle for the convex function $\phi(p)$, indicating the point $p^{(n)}$ results in the minimum of $\phi(p)$.
\section{Proof of Proposition 3}
In order to prove this property, we first introduce the following lemma:
\newtheorem*{lemma}{\bf{Lemma 1}}
\begin{lemma}
In the $n$th iteration of algorithm VII.1, for $p^{(n)}\in SOL(\mathcal{Q},F)$ (i.e., exact solution of inner loop) and $p_\delta^{(n)}\in SOL_{\delta_n}(\mathcal{Q},F_\tau)$ (i.e., inexact solution of inner loop) we have:
\begin{equation}
\label{trineq}
||p_\delta^{(n-1)}-p_\delta^{(n)}||^2+||p_\delta^{(n)}-p^{(n)}||^2\le||p_\delta^{(n-1)}-p^{(n)}||^2+2\frac{{\delta_n}}{\tau}
\end{equation}
and consequently:
\begin{equation}
\label{cor1}
||p_\delta^{(n)}-p^{(n)}||^2\le||p_\delta^{(n-1)}-p^{(n)}||^2+2\frac{{\delta_n}}{\tau}.
\end{equation}
\begin{proof}
~

Since $p^{(n)} \in SOL(\mathcal{Q},F)$, setting $y = p_\delta^{(n)}$ in $VI(\mathcal{Q},F)$ yields:
\begin{equation}
F(p^{(n)})^T(p_\delta^{(n)}-p^{(n)})\ge 0.
\end{equation}
$F(p)$ is monotone, so:
\begin{equation}
\label{mono2}
F(p_\delta^{(n)})^T(p^{(n)}-p_\delta^{(n)})\le 0.
\end{equation}
On the other hand $p_\delta^{(n)}\in SOL_{\delta_n}(\mathcal{Q}, F_\tau)$, so according to \eqref{soldelta}, by setting $y = p^{(n)}$ we have:
\begin{equation}
\label{soldelta2}
F_\tau(p_\delta^{(n)})^T(p^{(n)}-p_\delta^{(n)})\ge -\delta_n.
\end{equation}
Thus, we can be certain that at least:
\begin{equation}
\label{least}
\frac{1}{2}d^T_n.e_n = (p_\delta^{(n)}-p_\delta^{(n-1)})^T(p^{(n)}-p_\delta^{(n)})\ge-\frac{\delta_n}{\tau},
\end{equation}
Where $d_n = p_\delta^{(n)}-p_\delta^{(n-1)}$ and $e_n = p^{(n)}-p_\delta^{(n)}$. Using \eqref{least}, we have:
\begin{align}
 {\left|\left|p_\delta^{(n-1)}-p^{(n)}\right|\right|}^2 =  
{\left|\left|\left(p_\delta^{(n-1)}-p_\delta^{(n)}\right)-\left(p^{(n)}-p_\delta^{(n)}\right)\right|\right|}^2 =
\notag
\\
{\left|\left|p_\delta^{(n-1)}-p_\delta^{(n)}\right|\right|}^2+{\left|\left|p^{(n)}-p_\delta^{(n)}\right|\right|}^2 
+ 4 d_n^Te_n \ge~~~~~~~~~~~~~~
\notag
\\
 {\left|\left|p_\delta^{(n-1)}-p_\delta^{(n)}\right|\right|}^2+{\left|\left|p^{(n)}-p_\delta^{(n)}\right|\right|}^2 -\frac{2\delta_n}{\tau}~~~~~~~~~~~~~~~~~~~~
\end{align}
Comparing the most left hand side of the above inequality with the right hand side, we reach the inequality \eqref{trineq}.

Furthermore, rearranging the inequality \eqref{trineq} we have:
\begin{align}
\label{cor2}
||p_\delta^{(n-1)}-p_\delta^{(n)}||^2+||p_\delta^{(n)}-p^{(n)}||^2-||p_\delta^{(n-1)}-p^{(n)}||^2\le
\notag
\\
2\frac{{\delta_n}}{\tau}.
\end{align}
As $||p_\delta^{(n-1)}-p_\delta^{(n)}||^2$ is a positive value, we can deduce assuredly that
\begin{equation}
\label{lem1-2}
||p_\delta^{(n)}-p^{(n)}||^2-||p_\delta^{(n-1)}-p^{(n)}||^2\le2\frac{{\delta_n}}{\tau},
\end{equation}
which yields inequality \eqref{cor1}.
\end{proof}
\end{lemma}
In the following, we introduce another lemma:
\newtheorem*{lemma2}{\bf{Lemma 2}}
\begin{lemma2}\cite[Lemma 2.2]{moudafi}
Let $a_n$, $b_n$, and $c_n$ be real and positive sequences, such that $\sum_{n=0}^\infty c_n < \infty$ and $a_{n+1} \le a_n - b_n + c_n$. Then $a_n$ converges and $\sum_{n=0}^\infty b_n<\infty$.
\\
\qed
\end{lemma2}
Using lemma 2 in the inequality \eqref{cor2} (i.e., setting $a_n = ||p_\delta^{(n-1)} - p^{(n)}||$ (so $a_{n+1} = ||p_\delta^{(n)} - p^{(n)}||$ ), $c_n = \frac{\delta_n}{\tau}$, and
$b_n = ||p_\delta^{(n-1)}-p_\delta^{(n)}||^2$ we have:
\begin{equation}
\label{converged}
\lim_{n\rightarrow\infty}||p_\delta^{(n)}-p^{(n)}||^2<\infty.
\end{equation}
Using the result of \eqref{converged} in \eqref{trineq} and considering the assumptions made on $\frac{\delta_n}{\tau}$, we can conclude that
\begin{equation}
\lim_{n\rightarrow \infty}||p_\delta^{(n)}-p_\delta^{(n-1)}|| = 0.
\end{equation}
\section{Proof of Theorem 5}
This theorem states that with a lower number of iterations for the inner loop of algorithm \ref{algcontrol}, we can still reach to the minimum point of our merit function (i.e., $\phi(p)$).

In order to prove this theorem, we use the same reasoning as we presented in the proof of theorem 4. We let the solution set of \eqref{merit} as $\mathcal{S} \subset SOL(\mathcal{Q},F)$. While in the $n$th iteration, let us introduce the mapping $G_{\mathcal{S}}(p_\delta^{(n-1)})$ as the euclidean projection of the vector $p_\delta^{(n-1)}$ on the set $\mathcal{S}$. Going under the same procedure as in theorem 4, the following equations yield:
\begin{align}
\label{vimon}
F(G_{\mathcal{S}}(p_\delta^{(n-1)}))^T(p_\delta^{(n)}-G_{\mathcal{S}}(p_\delta^{(n-1)}))\ge 0,
\notag
\\
\\
F(p_\delta^{(n)})^T(G_{\mathcal{S}}(p_\delta^{(n-1)})-p_\delta^{(n)})\le 0.
\notag
\end{align}
Replacing $p_\delta^{(n)}$ in $VI(\mathcal{Q},F_{\epsilon, \tau})$ we have:
\begin{align}
\left[ F(p_\delta^{(n)})+\epsilon^{(n)}\nabla\phi(p_\delta^{(n)})\right]^T(y-p_\delta^{(n)})\ge 
\notag
\\
\tau(p_\delta^{(n-1)}-p_\delta^{(n)})(y-p_\delta^{(n)})-\delta_n.
\end{align}
The result of proposition 3 suggests that $\lim_{n\rightarrow \infty}||p_\delta^{(n)}-p_\delta^{(n-1)}|| = 0$. Hence, 
\begin{equation}
\left[ F(p_\delta^{(n)})+\epsilon^{(n)}\nabla\phi(p_\delta^{(n)})\right]^T(y-p_\delta^{(n)})\ge -\delta_n.
\end{equation}
setting $y = G_\mathcal{S}(p_\delta^{(n-1)})$ and considering \eqref{vimon}:
\begin{align}
\label{expand}
\delta_n \ge F(p_\delta^{(n)})^T(G_{\mathcal{S}}(p_\delta^{(n-1)})-p_\delta^{(n)})+\delta_n\ge 
\notag
\\
~
\\
\epsilon^{(n)}\nabla\phi(p_\delta^{(n)})^T(p_\delta^{(n)}-G_{\mathcal{S}}(p
_\delta^{(n-1)})),
\notag
\end{align}
We have assumed that $lim_{n\rightarrow \infty}\frac{\delta_n}{\tau} = 0$, so
\begin{align}
\label{mindelta}
\nabla\phi(p_\delta^{(n)})^T(G_{\mathcal{S}}(p_\delta
^{(n-1)})-p_\delta^{(n)})\ge 0.
\end{align}
Therefore, as $n \rightarrow \infty$, for all $p\in SOL_{\delta_n}(\mathcal{Q},F)$, the inequality in \eqref{expand} shows the minimum principle for the convex function $\phi(p)$, indicating that the point $p_\delta^{(n)}$ is the minimum point for $\phi(p)$.

\ifCLASSOPTIONcaptionsoff
  \newpage
\fi
\bibliographystyle{IEEEtran}
\bibliography{reff2}
\end{document}